\documentclass[twocolumn,aps, showpacs, showkeys]{revtex4}
\usepackage{graphicx}
\begin{document}

\title{Modulational instability criteria for two-component
 Bose--Einstein condensates \footnote{Submitted to
\textsl{European Physics Journal B}.}}

\author{
I. Kourakis$^{1, }$\footnote{Electronic address:
\texttt{ioannis@tp4.rub.de} ; \\
\texttt{http://www.tp4.rub.de/$\sim$ioannis}.}, P. K. Shukla$^{1,
2, }$\footnote{Electronic address: \texttt{ps@tp4.rub.de} ; \\
\texttt{http://www.tp4.rub.de/$\sim$ps}.}, M. Marklund$^{2}$ and L. Stenflo$^{2}$}
\affiliation{$^{1}$ Institut f\"ur Theoretische Physik IV,
Fakult\"at f\"ur Physik und Astronomie, \\
Ruhr--Universit\"at Bochum, D-44780 Bochum, Germany \\
$^{2}$ Department of Physics, Ume\aa \ University, SE-90187
Ume\aa,~Sweden}

\date{Submitted 13 April 2005; revised 12 May 2005}

\begin{abstract}
The stability of colliding Bose-Einstein condensates is investigated.
A set of coupled Gross-Pitaevskii equations is thus considered, and analyzed via a perturbative approach.
No assumption is made on the signs (or magnitudes) of the relevant parameters like the scattering lengths
and the coupling coefficients. The formalism is therefore valid for asymmetric as well as symmetric
coupled condensate wave states. A new set of explicit criteria is derived and analyzed. An extended
instability region, in addition to an enhanced instability growth rate is predicted
for unstable two component bosons, as compared to the individual (uncoupled) state.
\end{abstract}


\maketitle

\section{Introduction}
Bose-Einstein condensation of dilute gases in traps has attracted a great deal of interest
recently, as witnessed in recent reviews and monographs \cite{book1, book2}.
Mean-field theory provides a consistent framework for the modeling of the principal characteristics of
condensation and elucidates the role of the interactions between the particles. A generic theoretical
model widely employed involves the Gross-Pitaevskii equation, which bears the form of a
nonlinear Schr\"odinger-type equation, taking into account boson interactions
(related to a scattering length $a$), in addition to the confinement potential
imposed on the Bose-Einstein condensates (BECs) in a potential trap.
The scattering length $a$, although initially taken to be positive
(accounting for repulsive interactions and prescribing condensate stability), has later been
sign-inverted to negative (attractive interaction) via Feshbach resonance, in appropriately
designed experiments. This allowed for the prediction of BEC state instability, eventually
leading to wave collapse, which is only possible in the attractive case ($a < 0$) \cite{book1}.
As expected from previous know-how on problems modelled by generic nonlinear Schr\"odinger-type
equations (in one or more dimensions), the analysis of BEC dynamics revealed the possibility for
the existence of collective excitations including bright- (for $a < 0$) and dark- (holes, for $a > 0$)
type envelope excitations, as well as vortices, which were quite recently observed in
laboratories \cite{Myatt, Denschlag, Strecker, Khaykovich}. The evolution of coupled (``colliding'')
BEC wavepackets was recently considered in theoretical and experimental investigations
\cite{Timmermans, Heinzen, Band}. Pairs of nonlinearly coupled BECs are thus modeled via coupled
Gross-Pitaevskii equations, involving extra coupling terms whose sign and/or magnitude are a
priori not prescribed. Although theoretical modeling, quite naturally, first involved symmetric pairs
of (identical) BECs, for simplicity, evidence from experiments suggests that asymmetric boson pairs
deserve attention  \cite{Myatt}.

In this paper, we investigate the stability of a nonlinearly coupled BEC pair, from first principles.
Both BECs are assumed to lie in the ground state, for simplicity,  although no other assumption is made
on the sign and/or magnitude of relevant physical parameters. We shall derive a set of general criteria
for the stability of BEC pairs (allowing for asymmetry in the wave functions).

\section{The formalism}

The wave-functions $\psi_1$ and $\psi_2$ of two nonlinearly
interacting BECs evolve according to the coupled Gross-Pitaevskii
equations (CGPEs)
\begin{eqnarray}
i \hbar \frac{\partial \psi_1}{\partial t}  + \frac{\hbar^2}{2m_1}
\nabla^2 \psi_1 - V_{11} |\psi_1|^2\psi_1 - V_{12} |\psi_2|^2
\psi_1 \nonumber \\ + \mu_1 \psi_1
=0 \, , \nonumber \\
i \hbar \frac{\partial \psi_2}{\partial t}  + \frac{\hbar^2}{2m_2}
\nabla^2 \psi_2 - V_{22} |\psi_2|^2\psi_2 - V_{21} |\psi_1|^2
\psi_2 \nonumber \\ + \mu_2 \psi_2
=0 \, ,\nonumber \\
\label{CGPE}
\end{eqnarray}
where $\nabla^2 = \partial^2/\partial x^2 + \partial^2/\partial
y^2 + \partial^2/\partial z^2$ is the Laplace operator (a
three-dimensional Cartesian geometry is considered, for clarity).
Here $m_j$ represents the mass of the $j$th condensate. According
to standard theory, the nonlinearity coefficients $V_{jj}$ are
proportional to the scattering lengths $a_{j}$ via $V_{jj} =4 \pi
\hbar a_{j}/m_{j}$, while the coupling coefficients $V_{jl}$ are
related to the mutual interaction scattering lengths $a_{jl}$ via
$V_{jl} =2\pi\hbar a_{jl}/m_{jl}$, where $m_{jl}=m_j m_l/(m_j +
m_l)$ is the reduced mass. The (linear) last terms in each
equation involve the chemical potential $\mu_j$, which corresponds
to a ground state of the condensate, in a simplified model. These terms may readily be eliminated
via a simple phase-shift transformation, viz. $\psi_j = \psi'_j
\exp(i \mu_j t)$ ($j = 1, 2$); this is however deliberately not done at
this stage, for generality. Nevertheless, one therefore
intuitively expects no major influence of the chemical potentials
on the coupled BEC dynamics (at least for the physical problem studied
here).

\section{Linear stability analysis}

We shall seek an equilibrium state in the form $\psi_j = \psi_{j0}
\exp[i\varphi_j (t)]$, where $\psi_{j0}$ is a (constant real)
amplitude and $\varphi_j (t)$ is a (real) phase, into the CGP Eqs.
(\ref{CGPE}). We then find a monochromatic (fixed-frequency) Stokes'
wave solution in the form: $\varphi_j (t) = \Omega_{j0} t$, where
\[\Omega_{j0} = - \frac{V_{jj}}{\hbar} \psi_{j0}^2 -
\frac{V_{jl}}{\hbar} \psi_{l0}^2 + \mu_j \, , \] for $j \neq l=1,
2$.

Let us consider a small perturbation around the stationary state
defined above by taking $\psi_j =(\psi_{j0} + \epsilon
\psi_{j1})\exp[i\varphi_j (t)]$, where $\psi_{j1}({\bf r}, t)$ is
a complex number denoting the small ($\epsilon \ll 1$)
perturbation of the slowly varying modulated bosonic
wave-functions (it includes both amplitude and phase corrections),
and $\varphi_j (t)$ is the phasor defined above. Substituting into
Eqs. (\ref{CGPE}) and separating into real and imaginary parts by
writing $\psi_{j1} =a_j + i b_j$, the first order terms (in
$\epsilon$) yield
\begin{eqnarray}
- \hbar \frac{\partial b_j}{\partial t}  + \frac{\hbar^2}{2 m_j}
\nabla^2 a_j - 2 V_{jj}
\psi_{10}^2 a_j - 2 V_{jl} \psi_{j0} \psi_{l0} a_l =0 \, , \nonumber \\
\hbar \frac{\partial a_j}{\partial t}  + \frac{\hbar^2}{2m_j}
\nabla^2 b_j =0 \, , \quad \label{perteqs1}
\end{eqnarray}
where $j$ and $l (\neq j)=1, 2$ (this will be henceforth
understood unless otherwise stated). Eliminating $b_j$, these
equations yield
\begin{eqnarray}
\biggl[\hbar^2 \frac{\partial^2 }{\partial t^2} + \frac{\hbar^2}{2
m_1} \biggl(\frac{\hbar^2}{2 m_1} \nabla^2  - 2 V_{11}
\psi_{10}^2\biggr) \nabla^2\biggr] a_1 \nonumber \\
- \frac{\hbar^2}{m_1} V_{12} |\psi_{10}| |\psi_{20}| \nabla^2 a_2
= 0 \, , \quad \label{perteqs2}
\end{eqnarray}
(together with a symmetric equation, obtained by permuting $1
\leftrightarrow 2$). We now let $a_j = a_{j0}
\exp[i(\mathbf{k}\cdot \mathbf{r} - \Omega_k t)] + $ complex
conjugate, where ${\bf k}$ and $\Omega_k$ are the wavevector and
the frequency of the modulation, respectively, viz.
$\partial/\partial t \rightarrow - i \Omega_k$ and
$\partial/\partial x_n \rightarrow i k_n$ $(x_n \equiv \{ x, y, z
\}$ for $n = 1, 2, 3$) i.e. $\partial^2/\partial t^2 \rightarrow -
\Omega_k^2$ and $\nabla^2 \rightarrow - k^2$. After some algebra,
we obtain the eigenvalue problem: $\mathbf{M a} = (\hbar \omega)^2
\mathbf{a}$, where $\mathbf{a} = (a_1, a_2)^T$, and the matrix
elements are given by $M_{jj} = e_j(e_j +  2 V_{jj} |\psi_{j0}|^2)
\equiv \hbar^2 \Omega_{j}^2$ and $M_{jl} = - 2 e_j V_{jl}
|\psi_{j0}| |\psi_{l0}|\equiv \hbar^ 2 \Omega_{jl}^2$; where we
have defined $e_j = \hbar^2 k^2/2 m_j$. The frequency $\omega$ and
the wave number $\mathbf{k}$ are therefore related by the
dispersion relation
\begin{equation}
\left(\Omega_k^2 - \Omega_1^2\right)\left(\Omega_k^2 -
\Omega_2^2\right) = \Omega_{c}^4 \, , \label{DR}
\end{equation}
where the coupling is expressed via $\Omega_{c}^4 = \Omega_{12}^2
\Omega_{21}^2 \equiv M_{12} M_{21}/\hbar^4$ in the right-hand side
of Eq. (4). We stress that this dispersion relation
(which is independent of the chemical potentials $\mu_j$) relies
on absolutely {\em no} assumption on the sign or the magnitude of
$m_j$, $V_{jj}$ and $V_{jl}$.

\section{Modulational instability of individual BECs}

In the vanishing coupling limit, i.e. for $V_{jl} \rightarrow 0$,
the dispersion relation (\ref{DR}) gives $\Omega_{k, \pm} = \pm
\Omega_j$ ($j = 1, 2$). Absolute stability is ensured if $V_{jj} > 0$.
On the other hand, if $V_{jj} < 0$, a purely
growing unstable mode occurs (viz. $\Omega_k^2 < 0$) for
wavenumbers below a critical value $k_{j, cr}= 2 (m_j
|V_{jj}|)^{1/2} |\psi_{j0}|/\hbar$. The growth rate $\sigma = i
\sqrt{-\Omega_k^2}$ attains a maximum value $\sigma_{max} =
|V_{jj}| |\psi_{j0}|^2/\hbar$ at $k = k_{j, cr}/\sqrt{2}$.

Recalling the definitions of $V_{jj}$, we see that a
repulsive/attractive scattering length (i.e. positive/negative
$V_{jj}$) prescribes a stable/unstable (single) BEC behavior. In
the following, we shall see how this simple criterion for
stability ($V_{jj} > 0$) is modified by the presence of
interaction between two he condensates.

\section{Modulational instability of coupled BECs}

The dispersion relation (\ref{DR}) takes the form of a
bi-quadratic polynomial equation
\begin{equation}
\Omega_k^4 - T \Omega_k^2 + D = 0 \, , \label{DR2}
\end{equation}
where $T = {\rm{Tr}}\mathbf{M}/\hbar^2 \equiv \Omega_{1}^2 +
\Omega_{2}^2$ and $D = {\rm{Det}}\mathbf{M}/\hbar^4 \equiv
\Omega_{1}^2 \Omega_{2}^2 - \Omega_{12}^2 \Omega_{21}^2$ are
related to the \textsl{trace} and the \textsl{determinant},
respectively, of the matrix $\mathbf{M}$. Eq. (\ref{DR2}) has the
solution
\begin{equation}
\Omega_k^2 =\frac {1}{2} \bigl[ T \pm (T^2 -4  D)^{1/2} \bigr] \,
, \label{solution}
\end{equation}
or
\begin{equation}
\Omega_{k, \pm}^2 =\frac {1}{2} (\Omega_1^2+\Omega_2^2) \pm \frac
{1}{2} \left[ (\Omega_1^2 - \Omega_2^2)^2 + 4
\Omega_{c}^4\right]^{1/2} \, . \label{solution2}
\end{equation}
We note that the right-hand side is real/complex if the
\textsl{discriminant} quantity $\Delta = T^2 - 4 D$ is
positive/negative, respectively.

Stability is ensured (for any wavenumber $\mathbf{k}$) if (and
only if) \emph{both} solutions $\Omega_{k, \pm}^2$ are positive.
This is tantamount to the
following requirements being satisfied simultaneously: $T > 0$, $D
> 0$ and $\Delta > 0$. Since the three  quantities $T$, $D$ and
$\Delta$ are all even order polynomials of $k$, one has to
investigate three distinct polynomial inequalities. The stepstones
of the analysis will be outlined in the following, though trying
to  avoid burdening the presentation with unnecessary details.

First, the sign of $T = k^2 [(\hbar^2 k^2/4)
\sum_j (1/m_j^2) + \sum_j V_{jj} |\psi_{j0}|^2/m_j]$ (see
definitions above) depends on (the sign of) the quantity $\sum_j
V_{jj} |\psi_{j0}|^2/m_j$ which has to be positive for all $k$,
in order for stability to be ensured (for any $\psi_{j0}$ and
$k$). This requires that
\begin{equation}
V_{11} > 0 \quad and \quad  V_{22} > 0 \, . \label{C1}
\end{equation}
Otherwise, $T$ becomes negative (viz. $\Omega_{k, -}^2 < 0$, at
least) for $k$ below a critical value $k_{cr, 1} = \sqrt{K_1}$,
where $K_1 = 4 (- \sum_j V_{jj} |\psi_{j0}|^2/m_j)/[\hbar^2 \sum_j
({1}/{m_j^2})] > 0$ (cf. the single BEC criterion above); this is
always possible for a sufficiently large perturbation amplitude
$|\psi_{10}|$ if, say, $V_{11} < 0$ (even if $V_{22} > 0$).
Therefore, only a pair of two repulsive type BECs can be stable;
the presence of one attractive BEC may de-stabilize its
counterpart (even if the latter would be individually stable).

Second, $D = \Omega_{1}^2 \Omega_{2}^2 - \Omega_{12}^2
\Omega_{21}^2$ is an 8th-order polynomial in $k$, which can be
factorized as $D \sim k^4 (k^4 + b k^2 + c)$, where $b = 4 \sum_j
(m_j V_{jj})/\hbar^2$ and $c = 16 m_1 m_2 (V_{11}V_{22} - V_{12}
V_{21}) |\psi_{10}|^2 |\psi_{20}|^2/\hbar^4$ (note that $b^2 - 4 c
> 0$).
 The stability requirements $b > 0$ and $c > 0$ (in order for $D$
to be positive for \emph{any} value of $k > 0$) amount to $m_1
V_{11} + m_2 V_{22} > 0$ and
\begin{equation}
V_{11}V_{22} - V_{12} V_{21} > 0 \,  , \label{C2}
\end{equation}
respectively. Only the latter condition for stability has to be
retained, since the former one is automatically covered by
(8) above. To be specific, solving $D = 0$ for $k^2 =
K_{2, \pm}$, viz. $K_{2, \pm} = [- b \pm (b^2 - 4 c)^{1/2}]/2$, we
see that:\\ (i) if $b < 0 < c$, then $0 < K_{2, -} < K_{2, +}$,
and $D < 0$ for $\sqrt{K_{2, -}} < k < \sqrt{K_{2, +}}$
(instability for short wavelengths);\\ (ii) if $c < 0$ (regardless
of $b$), then $K_{2, -} < 0 < K_{2, +}$, and $D < 0$ for $0 < k <
\sqrt{K_{2, +}}$;
\\(iii) if $b > 0$ and $c > 0$, then $K_{2, -} < K_{2, +} < 0$, so that $D > 0$.
\\We see that this kind of instability,
-- i.e. if the criterion (9) is not met, is due to the mutual
interaction potential $V_{jl}$ among the bosons.

Finally, the positivity of $\Delta = T^2 - 4 D = (\Omega_1^2 -
\Omega_2^2)^2 + 4 \Omega_{12}^2 \Omega_{21}^2$ is only ensured
(for every value of $k$ and $|\psi_{j0}|$) if $\Omega_{12}^2
\Omega_{21}^2 \sim M_{12} M_{21}
> 0$, i.e. if
\begin{equation}
V_{12} V_{21} > 0 \,  .
\label{C3}
\end{equation}
If this condition is not met, the solution (\ref{solution}) above
has a finite imaginary part, which accounts for amplitude
instability due to the external perturbation. For rigour, we note that
$\Delta$ bears the form $\Delta = k^4 (c_4 k^4 - c_2 k^2 + c_0)$
(where $c_4 > 0$; the complex expressions for $c_n$ are omitted).
If $\Delta' \equiv c_2^2 - 4 c_0 c_4 \sim - V_{12} V_{21} < 0$,
i.e. if (10) is met, then $\Delta > 0$ for any value of
$k$. If $\Delta' > 0$, on the other hand, denoting $K_{3, \pm} =
[c_2 \pm (c_2^2 - 4 c_0 c_4)^{1/2}]/(2 c_4)$, we find that: \\
(i) stability is only ensured (since $K_{3, -} < K_{3, +} < 0 <
k^2$)
 if $c_2 < 0 < c_0$ (nevertheless, this condition depends on the
perturbation amplitudes $|\psi_{j0}|$ and may always be violated).
\\
(ii) Again, a finite unstable wavenumber interval $k \in
(\sqrt{K_{3, -}}, \sqrt{K_{3, +}})$ is obtained for $c_2 > 0$ and
$c_0 > 0$. \\
(iii) Finally, instability will be observed for $k \in (0,
\sqrt{K_{3, +}})$ if $c_0 < 0$ (regardless of $c_2$).

\section{Conclusions}

Summarizing, we have derived a set of explicit criteria, (8) to
(10) above, which should \emph{all} be satisfied in order for a
boson pair to be stable. Therefore, an interacting BEC pair is
\emph{stable} only if the interaction potentials satisfy $V_{11} >
0$ and $V_{22} > 0$ and $V_{11} V_{22} > V_{12} V_{21} > 0$. If
one criterion is not met, then the perturbation frequency develops
a finite imaginary part and the solution blows up in time. A few
comments and qualitative conclusions should however be mentioned.

\begin{figure}[floatfix]
\includegraphics[width=3.0in]{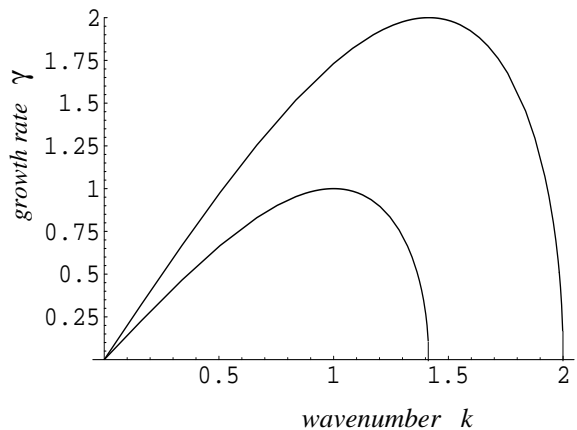}
\caption{The growth rate $\gamma$ versus the wavenumber $k$ (in
units of $|V_{jj}| |\psi_{j, 0}|^2/\hbar$ and $\sqrt{2 m_j
|V_{jj}|} |\psi_{j, 0}|/\hbar$, respectively) for a symmetric pair
of coupled unstable ($V_{jl} < 0$ for $j, l = 1, 2$) BECs (upper
curve) as compared to the single BEC case (lower curve).}
\label{figure1}
\end{figure}

First, for a symmetric stable boson pair, viz. $V_{11} = V_{22} >
0$ and $V_{12} = V_{21}$, stability is ensured if $V_{12}^2 <
V_{11}^2$. Second, if \emph{one} BEC satisfies $V_{jj} < 0$, the
pair will be unstable: only pairs consisting of stable bosons can
be stable. Interestingly, in the case of individually unstable
BECs (viz. $V_{jj} < 0$, for $j=$1 or 2), the instability
characteristics are strongly modified. For instance, in the case
of a symmetric unstable boson pair (viz. $V_{11} = V_{22} < 0$ and
$m_1 = m_2$), an \emph{extended} unstable wavenumber region and an
enhanced growth rate can be obtained, as can be checked via a
tedious calculation; cf. Fig. \ref{figure1}. Furthermore, we have
pointed out the appearance of secondary instability ``windows'',
i.e. unstable wave number intervals beyond $(k_{cr}, k'_{cr})$,
where $k_{cr} \ne 0$.

These results above follow from a set of explicit stability criteria.
Both BECs were assumed to lie in the ground state, for simplicity,
although no other assumption was made on the sign and/or magnitude of the relevant physical parameters.
Naturally, a future extension of this work should consider the external confinement potential,
imposed on the trapped condensates.
Our results can be tested, and can hopefully be confirmed, by designed experiments.


\begin{acknowledgments}
I.K. is grateful to the Max-Planck-Institut f\"ur
extraterrestrische Physik (Garching, Germany) for the award of a
fellowship (project: Complex Plasmas). Partial
support from the Deutsche Forschungsgemeinschaft through
the Sonderforschungsbereich (SFB) 591 -- Universelles Verhalten
Gleichgewichtsferner Plasmen: Heizung, Transport und
Strukturbildung is also gratefully acknowledged.
\end{acknowledgments}


\end{document}